\documentclass[twocolumn]{aastex61}
\usepackage{hyperref} 
\usepackage{xspace,graphics,graphicx,longtable,amsmath}
\usepackage{natbib,amsbsy}
\usepackage{appendix}
\usepackage{natbib}

\providecommand{\zabs}{\ensuremath{z_{\rm abs}}}
\providecommand{\rAA}{\ensuremath{\,\mbox{\AA}}}

\providecommand{\cm[1]}{\ensuremath{\,{\rm cm}^{#1}}} % KLC: \, adds nice little space

\providecommand{\kms}{\ensuremath{\,{\rm km\,s}^{-1}}}
\providecommand{\Lya}{\ensuremath{{\rm Ly}\alpha}}
\providecommand{\Lyaf}{\Lya\ forest}

\providecommand{\N[1]}{\ensuremath{N_{\rm #1}}}

\providecommand{\om}{\ensuremath{\Omega_{\rm M}}}

\providecommand{\EWr}{\ensuremath{W_{\rm r}}}
\providecommand{\sigEWr}{\ensuremath{\sigma_{\EWr}}}
\providecommand{\EWlin[1]}{\ensuremath{\EWr{}_{,\mathrm{#1}}}}

\providecommand{\EWo}{\ensuremath{W_{\rm obs}}}

\providecommand{\dv[1]}{\ensuremath{\delta v_{#1}}}

\def \object {J014709+463037}
\def \nickname {Andromeda's Parachute}
\providecommand{\zqso}{\ensuremath{2.377}}
\providecommand{\zqsoerr}{\ensuremath{0.007}}

\shorttitle{Quadruply Lensed Quasar}
\shortauthors{Rubin et al.}
\submitjournal{ApJ Letters}
\begin{document}
\title{Andromeda's Parachute: A Bright Quadruply Lensed Quasar at  \lowercase{$z$}$=\zqso$}

\author{Kate~H.~R.~Rubin}
\affiliation{Department of Astronomy, San Diego State University, 
  San Diego, CA 92182, USA} \author{John~M.~O'Meara}
\affiliation{Department of Physics, Saint Michael's College, One Winooski Park, Colchester, VT 05439} \author{Kathy~L.~Cooksey}
\affiliation{Department of Physics \& Astronomy, University of
  Hawai`i at Hilo, 200 West K\=awili Street, Hilo, HI 96720, USA} 
\author{Mateusz Matuszewski}
\affiliation{Cahill Center for Astrophysics, California Institute of Technology, 1216 East California Boulevard, Mail Code 278-17, Pasadena, California 91125, USA}
\author{Luca Rizzi}
\affiliation{W.M. Keck Observatory, 65-1120 Mamaloha Hwy,
Kamuela, HI 96743} 
\author{Gregory W.~Doppmann}
\affiliation{W.M. Keck Observatory, 65-1120 Mamaloha Hwy,
Kamuela, HI 96743} 
\author{Shui Hung Kwok}
\affiliation{W.M. Keck Observatory, 65-1120 Mamaloha Hwy,
Kamuela, HI 96743} 
\author{D. Christopher Martin}
\affiliation{Cahill Center for Astrophysics, California Institute of
  Technology, 1216 East California Boulevard, Mail Code 278-17,
  Pasadena, California 91125, USA}
\author{Anna M. Moore}
\affiliation{Research School of Astronomy and Astrophysics,
Australian National University, Canberra, ACT 2611, Australia}
\author{Patrick Morrissey} 
\affiliation{Cahill Center for Astrophysics, California Institute of Technology, 1216 East California Boulevard, Mail Code 278-17, Pasadena, California 91125, USA}
\author{James D. Neill}
\affiliation{Cahill Center for Astrophysics, California Institute of Technology, 1216 East California Boulevard, Mail Code 278-17, Pasadena, California 91125, USA}

%\altaffiliation{1}{San Diego State University, Department of Astronomy,
%  San Diego, CA 92182, USA}
%\altaffiliation{2}{Department of Physics, Saint Michael's College, One Winooski Park, Colchester, VT 05439}
%\altaffiliation{3}{Department of Physics \& Astronomy, University of
%  Hawai`i at Hilo, 200 West K\=awili Street, Hilo, HI 96720, USA;
% kcooksey@hawaii.edu}
%\altaffiliation{4}{Cahill Center for Astrophysics, California Institute of Technology, 1216 East California Boulevard, Mail Code 278-17, Pasadena, California 91125, USA}
%\altaffiliation{5}{W.M. Keck Observatory, 65-1120 Mamalahoa Hwy.
%Kamuela, HI 96743}

%\author{D. Christopher Martin, Patrick Morrissey}
%\affil{Cahill Center for Astrophysics, California Institute of Technology, 1216 East California Boulevard, Mail Code 278-17, Pasadena, California 91125, USA }

%\author{James D. Neill}
%\affil{Cahill Center for Astrophysics, California Institute of Technology, 1216 East California Boulevard, Mail Code 278-17, Pasadena, California 91125, USA }

\correspondingauthor{Kate H. R. Rubin}
\email{krubin@sdsu.edu}

%\slugcomment{Draft 1: \today}

\begin{abstract}
We present Keck Cosmic Web Imager spectroscopy of the four putative images of the lensed quasar candidate \object\ recently discovered by \citet{bergheaetal17ph}.  The data verify the source as a quadruply lensed, broad absorption-line quasar having $z_{\rm S} = \zqso\pm\zqsoerr$.  
We detect intervening absorption in the \ion{Fe}{2} $\lambda\lambda 2586, 2600$, \ion{Mg}{2} $\lambda\lambda 2796, 2803$, and/or \ion{C}{4} $\lambda\lambda 1548, 1550$ transitions in eight foreground systems, three of which have redshifts consistent with the photometric-redshift estimate reported for the lensing galaxy ($z_{\rm L} \approx 0.57$).
By virtue of their positions on the sky, 
the source images probe these absorbers over transverse physical scales of $\approx\!0.3$--21\,kpc, permitting assessment of the variation in metal-line equivalent width \EWr\ as a function of sight-line separation.  We measure differences in $\EWlin{2796}$ of $<\!40\%$ across all sight-line pairs subtending 7--21\,kpc, suggestive of a high degree of spatial coherence for \ion{Mg}{2}-absorbing material.
$\EWlin{2600}$ is observed to vary by $>50$\% over the same scales
across the majority of sight-line pairs, while \ion{C}{4} absorption exhibits a wide range in $\EWlin{1548}$ differences of $\approx\!5$--80\% within transverse distances $\lesssim3\,$kpc.
\object\ is one of only a handful of $z > 2$ quadruply lensed systems 
for which all four source images are very bright ($r = 15.4$--17.7\,mag)
and are easily separated in ground-based seeing conditions.  As such, it is an ideal candidate for higher-resolution spectroscopy  probing the spatial variation in %the detailed velocity structure and ionization conditions of intervening absorbers. 
the kinematic structure and physical state of intervening absorbers.
\end{abstract}


\section{Introduction}\label{sec.intro}

Strong gravitational lensing of high-redshift quasars has proven a
powerful astrophysical and cosmological 
tool for a myriad of applications.  %probing phenomena over a
%great range of scales. 
Experiments range from high-fidelity spectroscopy probing 
the structure of the broad-line region
surrounding the host active galactic nuclei
\citep[e.g.,][]{rauchblandford91,sluse12}
%, to the
%coherence scale of diffuse intergalactic baryons
%\citep[e.g.,][]{young81,rauch99,chen14},  
to time domain observations constraining 
cosmological parameters %of the Big-Bang cosmological model
\citep[e.g.,][]{bonvin17}.
%high-fidelity spectroscopy 
%or time-domain observations of such systems have the potential to advance a variety of applications.
However, the brightest and most valuable of these sources are rare. 
%on the sky.  
%With the advent of 
Candidate lensed quasars 
may now be efficiently identified via color and morphological selection
techniques \citep[e.g.,][]{schechter17} or using variability criteria
\citep[e.g.,][]{kochanek06}
in wide-field optical and near-infrared imaging surveys 
\citep[e.g.,][]{inada12,diehl14,shanks15}.
%(e.g., SDSS, ATLAS, Pan-STARRS, DES).  
Follow-up spectroscopy is then always required to confirm the nature of the system.
 
%Spectroscopic follow-up, often at high resolution, is then required for most scientific programs.

Recently, \citet{bergheaetal17ph} identified a quadruply lensed quasar
candidate in imaging obtained by the Panoramic Survey Telescope and Rapid Response System \citep[hereafter PS1;][]{chambersetal16ph}. 
Astrometry of the components implies distances between source images of $\approx1.3$--$3.4\arcsec$. 
%Using spectral energy distribution (SED) modeling, 
They reported
satisfactory spectral energy distribution (SED) fits to the source photometry 
for quasar templates at both $z_{\rm S}=0.820^{+0.006}_{-0.007}$ and $z_{\rm S}\approx2.6$.
%Adopting an early-type galaxy template for analysis of the prospective
%lens galaxy SED, 
Additionally, they 
found that SED modeling of the photometry of the prospective lens galaxy yields
a best-fit redshift $z_{\rm L}=0.57^{+0.20}_{-0.13}$.

In principle, images of a source QSO at $z_{\rm S}\approx 2.6$ 
lensed by a foreground system at $z_{\rm L}= 0.57$ 
and separated by 1.3--$3.4\arcsec$
probe physical scales of $\approx\!0.5$--21\,kpc at $z\approx0.5$--2.
Such a configuration is highly valuable for study of the transverse
small-scale coherence of circumgalactic medium (CGM) absorption.
%Moreover, background light sources at $z\approx2.5$ typically probe numerous intervening gaseous structures along the line of sight.  Follow-up spectroscopic surveys of the surrounding fields may then efficiently identify the galaxy counterparts of these absorbers \citep[e.g.,][]{rudie12,crightonetal15}. %\textbf{(KLC: whoever had steidelxx and crightonxx, check I got what you wanted.)}  
%The images of this particular lensed quasar candidate are
%exceptionally bright ($r=15.4$--17.7\,mag), are easily separated in
%ground-based seeing conditions, and are not significantly affected by
%continuum emission from the putative lens.  
The brightness of this particular candidate  ($r=15.4$--17.7\,mag) and 
relatively large separation of the source images
enable high-signal-to-noise (S/N) spectroscopy %at high spectral resolution 
with maximum efficiency.  
% \textbf{Might be interesting to comment on whether these properties are also good for BLR, dark matter fraction, $H_0$.  Might also be interesting to know how many other quads there are at $z_{\rm S}>2$.  ALSO, might be interesting to know how rare such bright, high-z quads are in general.  LSST is supposed to discover 8000 lensed quasars \citep{oguri10}, but how many of them are going to be this bright??}

%estimation of cosmological parameters via analysis of the time delay between the source image light curves (XX, XX)

%enriched a wide variety of astrophysical applications since the initial discovery of this %phenomenon by XXX.  From constraints on the structure of the broad line region associated with the source (XX,XX) to estimation of cosmological parameters via analysis of the time delay between the source image light curves (XX, XX), strongly lensed QSOs probe a great range of physical scales. [This probably sucks so far...]

%However, they are unfortunately very rare, difficult to identify in imaging surveys (must be done by eye), and require spectroscopic confirmation...which can be expensive/awkward with single-slit spectrographs.

%% Figure 1: KCWI footprint --- KLC moved here because really wanted on p2
\begin{figure*}[!thb]
 \begin{center}
   \includegraphics[width=0.49\textwidth,clip=true,trim=1.5cm 0 2cm 0]{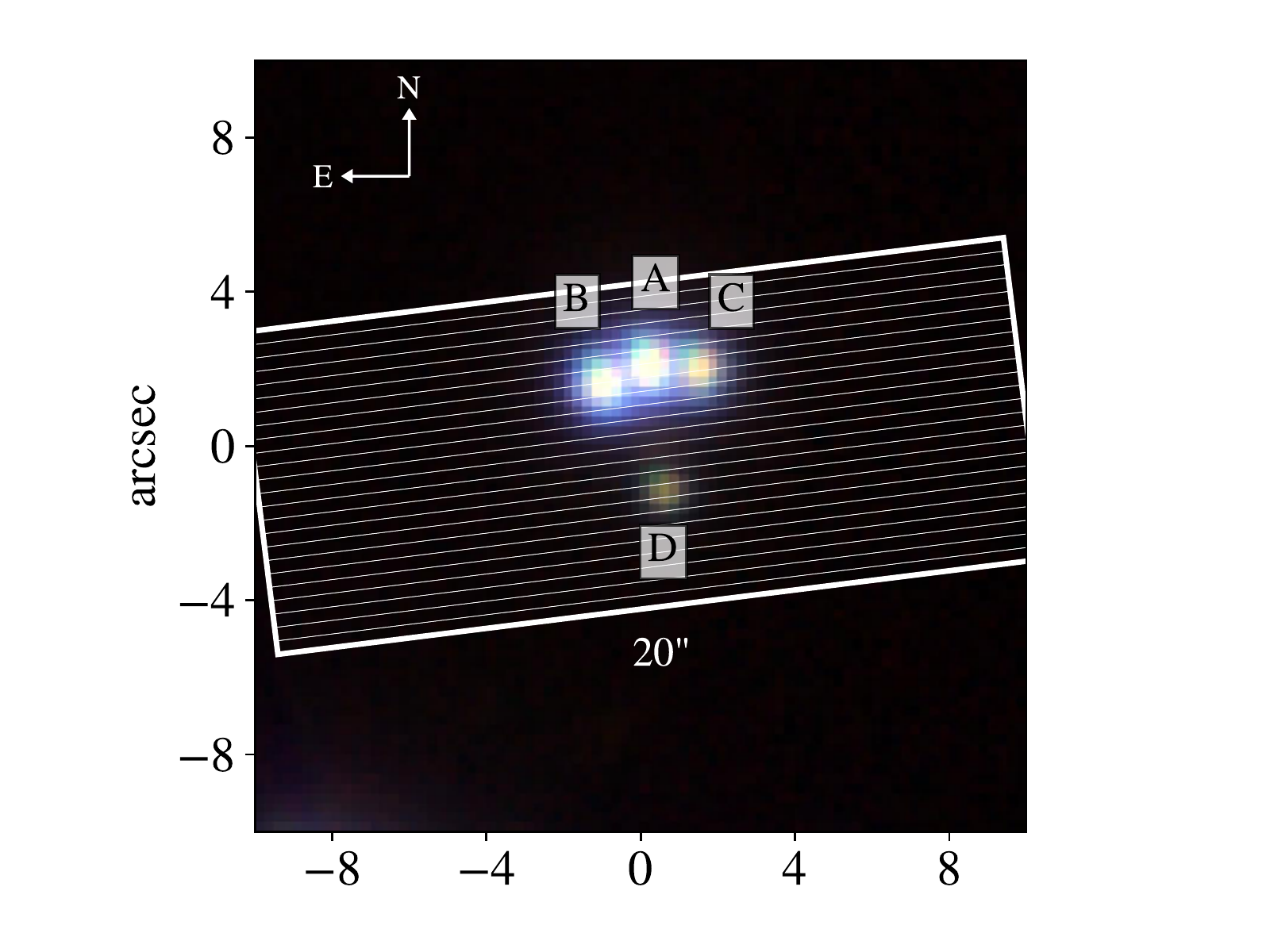}
   \includegraphics[width=0.39\textwidth,clip=true,trim=6.0cm 0 0cm 0]{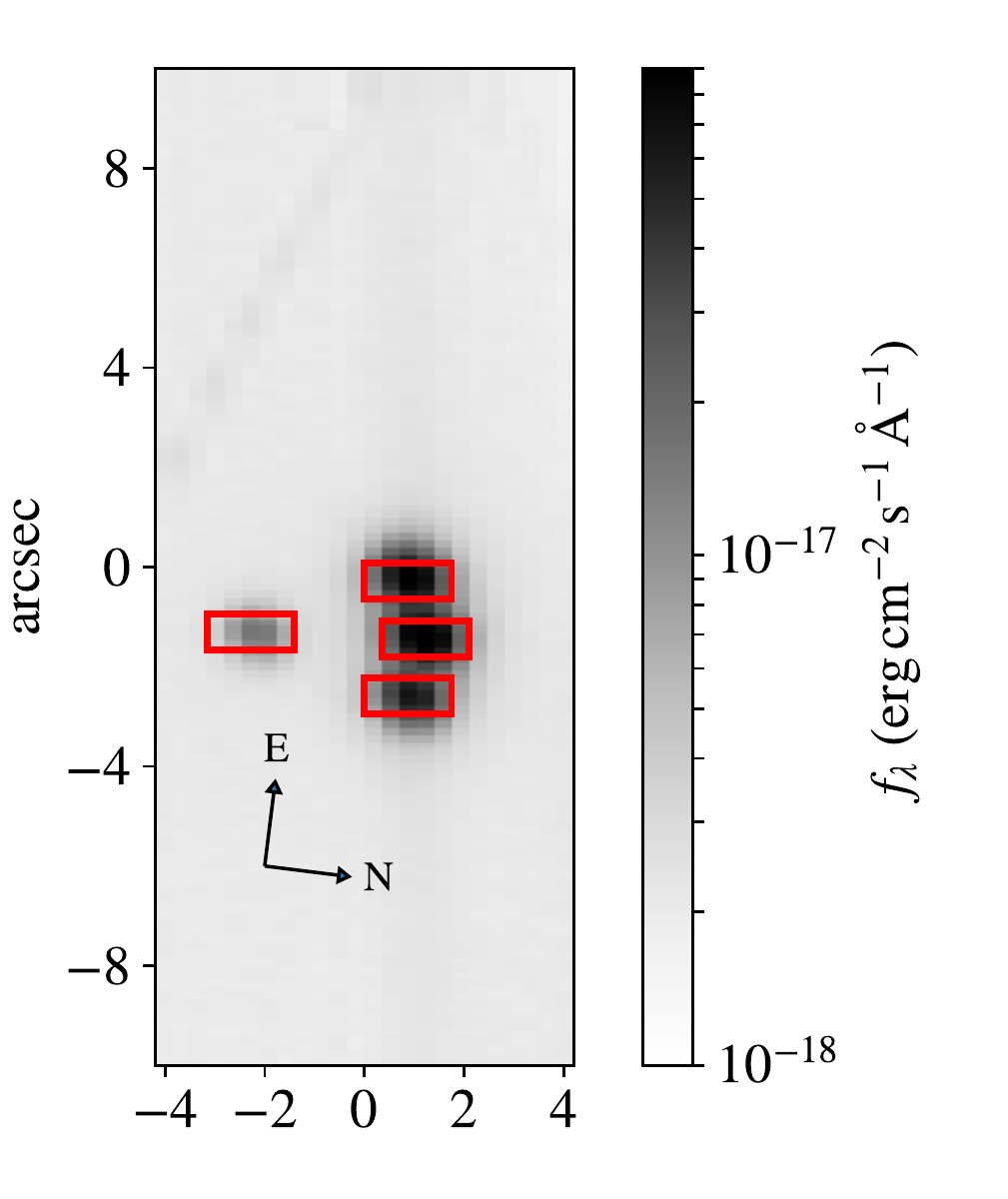}
 \end{center}
\vspace{-0.7cm}
 \caption%[KCWI footprint]
 {\emph{Left:} PS1 $giy$ color image  \citep{chambersetal16ph} of the
   lensed quasar system. Each source image  is labeled A--D 
(brightest to faintest) as in \citet{bergheaetal17ph}.  The thick
   white outline indicates the placement of the KCWI
   $20\arcsec\!\times\!8.4\arcsec$ field-of-view over the target.  The thin
   white lines show the width of each $0.35\arcsec$-image slicer
   within the KCWI footprint.
 \emph{Right:} KCWI image of the target obtained from one 600-s
 exposure.  The grayscale shows the flux density averaged over
 the wavelength range $4200\rAA\ <\lambda_{\rm obs}<4300\rAA$ in each
 spaxel of the rectified datacube.  The red boxes indicate the
 placement and size of the apertures used to generate the
 one-dimensional spectra shown in Figure~\ref{fig.kcwi1d}.
%% KHRR: seems like there's a weird artifact in the cube toward the upper left.  I wonder what that's about?
   \label{fig.kcwifoot}
 }
\end{figure*}

%\citet{bergheaetal17ph} state: ``To our knowledge, if proven with spectroscopic observations, [\object] would be the first gravitational lens discovered in the PS1 data.'' 
In this Letter, we present 
spectroscopy from the recently-commissioned 
Keck Cosmic Web Imager (KCWI; Morrissey, P. et al.\ 2017, in prep.)
%KCWI spectroscopy 
confirming that this
system (\object) is a quadruply lensed quasar at $z_{\rm S}=\zqso$.  
We then analyze intervening metal-line absorption systems for
constraints on their spatial coherence.
%use these spectra to 
%present analysis of 
%the spatial coherence of intervening metal-line absorption systems. %for constraints
%on their spatial coherence.  %detected along
%the sight lines.
 %As \citet{bergheaetal17ph} discuss, this object is likely the first gravitational lens discovered in PS1 data. 
Given the source image configuration and its location on the sky, we
refer to this object as ``\nickname.''\footnote{The NASA Apollo and
  Orion command modules and SpaceX Dragon capsule (source image D)
  made ocean landings with the aid of a parachute system (A--C).}
%As \citeauthor{bergheaetal17ph} note, only about three dozen quadruply lensed quasars are known so our confirmation increases the sample by $2.\bar{7}\%$. 
%We summarize our observations in Section \ref{sec.data}. 
%Redshift estimation and analysis of intervening absorption-line system 
%are described in Section \ref{sec.analysis}. We discuss the implications
%of our results
%for the transverse coherence scale of CGM absorption %and the results from surveys
%of other absorber populations 
%in Section \ref{sec.discuss}. 
We adopt the WMAP5
cosmology ($H_0=71.9\kms\,{\rm Mpc}^{-1}$, 
$\om=0.258$, and
$\Omega_{\Lambda}=0.742$; \citealt{komatsuetal09}) throughout this work unless otherwise specified.

%% Figure 2: Four sight lines
\begin{figure*}[hbt]
 \begin{center}
  \includegraphics[width=\textwidth,clip=true,trim=-0.1cm 0cm 0cm 0]{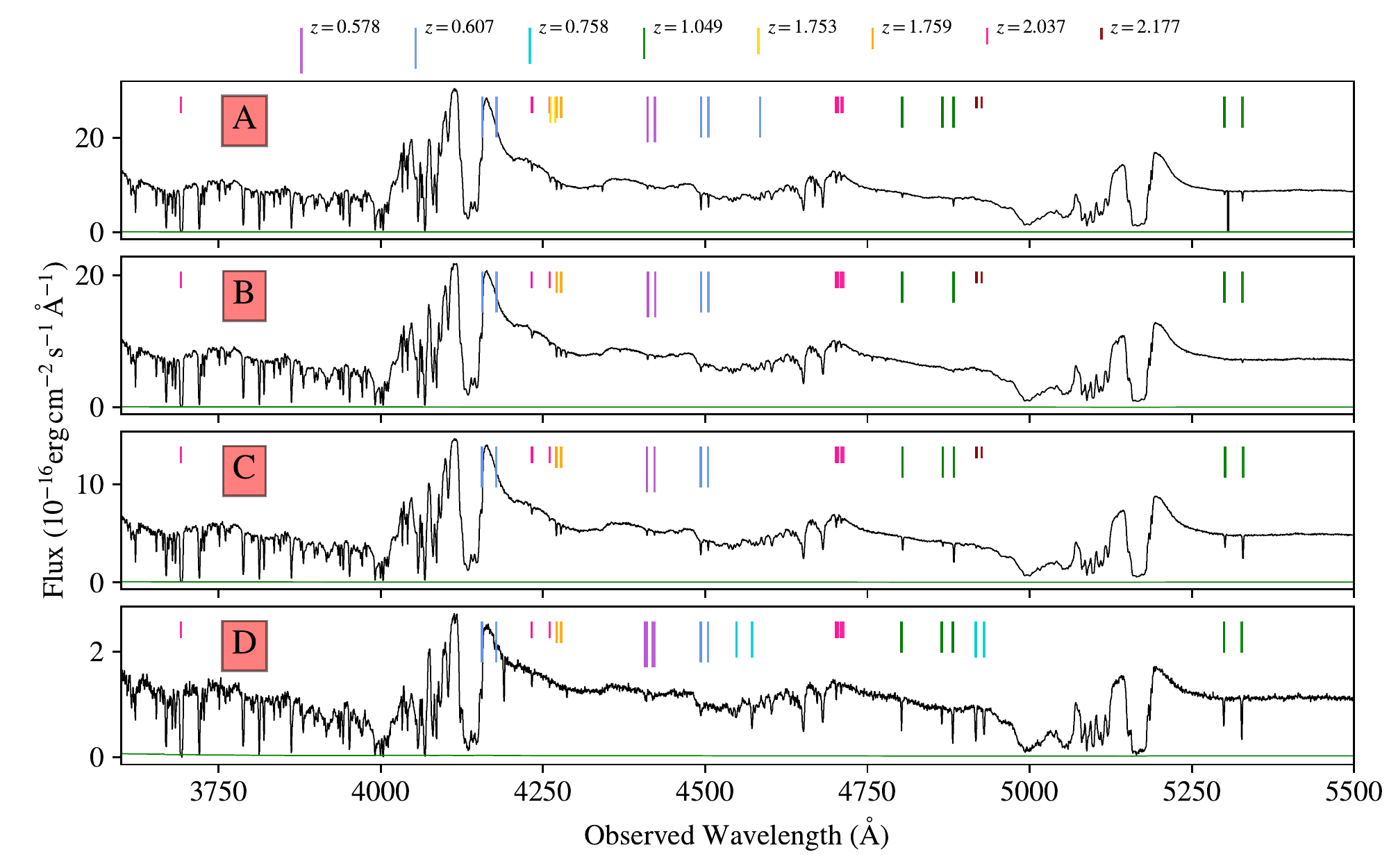}
  \end{center}
\vspace{-0.5cm}
  \caption%[Full 1D KCWI spectra]
 {Extracted 1D spectra of the four images indicated in Figure~\ref{fig.kcwifoot} demonstrating that the source
   is a lensed BAL quasar at $z_{\rm S}=\zqso$.  
   Intrinsic absorption features extend up to $\approx\!10,000\kms$
   blueward of each emission line (e.g., \ion{C}{4} at 
$\lambda_{\rm obs}\approx5200\rAA$). Intervening metal-line absorption
   systems are indicated with vertical colored marks, and are in most
   cases detected along all four sightlines.
   %The metal-line systems clearly show variation across sight lines (e.g., $\zabs = 1.049$ \ion{Fe}{2} lines at $\lambda_{\rm obs} \approx 5300\rAA$), which enable us to study the transverse size scales of the absorbing gas clouds. %In addition, there are velocity shifts between lines of sight, due to different kinematics (e.g., rotation). 
\label{fig.kcwi1d}
 }
\end{figure*}

\section{Observations}\label{sec.data}

%The recently-commissioned Keck Cosmic Web Imager (KCWI; Morrissey, P. et al. 2017, in prep.)
%\citep[KCWI;][]{martinetal10,kcwi} 
%KCWI is a seeing-limited integral field spectrograph on the 10-m Keck II Telescope. 
%Its configuration as an image slicer with a field of view spanning
%$>8\arcsec$ permits simultaneous spectroscopy of all four source
%images in this lensed QSO candidate in a single pointing without slit
%losses.  Its exquisite sensitivity at wavelengths $\lambda_{\rm obs}<5000~\mathrm{\AA}$ 
%is ideal for study of absorption in foreground \Lya\ and near-UV metal-line transitions.
%KCWI is an ideal instrument for \object\ follow-up, as its field of view covers all four images in a single pointing, there are no slit losses, and the high-sensitivity permits science-quality spectra with high-S/N coverage of the \Lya\ forest in a 10-min confirmation exposure. 

We observed  \object\ with KCWI on the night of 2017 June 21 UT.  
The instrument was configured with the Small image slicer and the BL
grating, providing a spatial sampling of $0.35\arcsec\,{\rm pix}^{-1}$ and
a spectral resolution $\mathcal{R}\approx5000$.
The field-of-view of KCWI in this configuration is 
$8.4\arcsec\times20\arcsec$,
%field-of-view 
permitting simultaneous spectroscopy of all four source
images in a single pointing. 
%This setup provides a spatial sampling of $0.35\arcsec\,{\rm pix}^{-1}$ and
%a spectral resolution $\mathcal{R}\approx5000$. 
%(full-width at half-maximum ${\rm FWHM} \approx 60\kms$). 
PS1 imaging of the target and the placement of  
the KCWI footprint is shown in Figure \ref{fig.kcwifoot} (left panel).
The  image slices were oriented at a position angle of $97.1^{\circ}$.
% (i.e., close to the parallactic angle $\approx\!77^{\circ}$). 
%the four lensed quasar images are labelled A--D in order of decreasing brightness. 
We obtained two exposures of 600\,s each 
while the target was at airmass $\approx1.5$--1.6.  

The data were reduced using the publicly-available \texttt{kderp}
package.\footnote{See \url{https://github.com/kcwidev/kderp}.}  
We used in-house software to rectify the curved object
traces along the cube resulting from differential atmospheric dispersion.
%The pipeline yields datacubes in which the spatial location of each
%source varies significantly as a function of wavelength due to
%differential atmospheric dispersion.  To rectify the object traces
%along the cube, we use in-house software to perform a non-linear
%least-squares fit of a two-dimensional Gaussian (using routines
%available in the \texttt{MPFIT} package; \citealt{markwardt09}) to
%one of the source images (D) at each spectral pixel. The best-fit
%spatial location of the source varies by $\approx\!1.1\,$pix in the
%$x$-direction and $\approx\!8\,$pix in the $y$-direction over the
%full spectral range of the cube.  Second-order polynomials are then
%fit to the resulting array of Gaussian centroids to smooth over any
%fitting-by-wavelength anomalies.  Finally, at each spectral pixel
%location, we shift the cube in the $x$- and $y$-directions by an
%amount specified by our polynomial fits.  
The final rectified KCWI image, averaged over the wavelength range
$4200~\rAA\ <\lambda_{\rm obs}<4300~\rAA$, is shown in the right-hand
panel of Figure~\ref{fig.kcwifoot}.

To extract one-dimensional spectra from this rectified cube, 
we performed a simple sum of the flux over the spatial dimensions 
in $5\!\times\!5$-pixel sub-cubes centered on each source image (see
red apertures in Figure \ref{fig.kcwifoot}, right panel).
A sky spectrum was extracted in the same manner from an off-source
region of the datacube and was subtracted from each of the on-source
spectra. We then co-added the 1D spectra extracted from the two 600-s
exposures for each source image.
%of each source image from the two 600-s exposures.  
These co-added 1D spectra are shown in Figure~\ref{fig.kcwi1d}, and have
median S/N per pixel measured redward of the quasar \Lya\ emission
line of $\sim180$, $\sim160$, $\sim130$, and $\sim45$ in images A,
B, C, and D, respectively.

\section{Spectroscopic Analysis}\label{sec.analysis}

%The striking similarity of the 1D spectra in 
Figure~\ref{fig.kcwi1d} makes evident that each source image
originates from the same high-redshift quasar.  %Moreover, 
The complex
absorption features blanketing the quasar's broad \Lya, \ion{Si}{4},
and \ion{C}{4} emission lines (at $\lambda_{\rm obs}\sim4100\,$\AA,
$4700\,$\AA, and $5150\,$\AA) indicate it belongs to the broad absorption
line (BAL) quasar subclass  \citep{weymann91}.  %,trump06}.  
%Having
%confirmed the nature of the source, we now describe our method of
%redshift estimation and analysis of intervening absorption systems.

%The KCWI data not only provide a tight constraint on the redshift of the quadruply-lensed background quasar, but are also of sufficient quality to reveal several foreground intervening absorption systems along the lines of sight. Here 

\subsection{Redshift of \object}\label{subsec.redshift}

%% KLC:
%%IDL>calc_zqso,/clobber
%%  running again on Coadded spectra
%% quadlens_kcwiall_1d_A.fits       2.38447    0.0082534625      2.38410
%% quadlens_kcwiall_1d_B.fits       2.38555    0.0045273373      2.38500
%% quadlens_kcwiall_1d_C.fits       2.38524    0.0016492693      2.38500
%% quadlens_kcwiall_1d_D.fits       2.38666    0.0013711311      2.38710
%%    <zqso> =       2.38548+/-  0.000454919; sigma =   0.000909838; max-min =    0.00219417
%% the +/- is stderr on mean: stddev / sqrt(N)

%%IDL>calc_zqso,wvnrm=1250.,/clobber
%% quadlens_kcwiall_1d_A.fits       2.37125   0.00081686976      2.37090
%% quadlens_kcwiall_1d_B.fits       2.37522    0.0032201369      2.37490
%% quadlens_kcwiall_1d_C.fits       2.37462    0.0024145752      2.37410
%% quadlens_kcwiall_1d_D.fits       2.38680    0.0012296701      2.38650
%%    <zqso> =       2.37697+/-   0.00338952; sigma =    0.00677904; max-min =     0.0155492

%% We can make all sorts of caveats of how the template doesn't look 
%% awesome but the quad-lensed quasar is a BAL and all quasars have funny 
%% emission/unique lines

%% FYI, template min rest wavelength is ~1190 Ang
To measure the quasar  redshift, we cross-correlate the quasar
template used in \citet{hewettandwild10} with each of the four 
spectra shown in Figure~\ref{fig.kcwi1d}.  %each
%source image.
%over a range of redshifts. 
Given the BAL nature of \object, we exclude
the template blueward of restframe $\lambda=1250\,$\AA\ (i.e., the
\Lya\ and \ion{N}{5} QSO emission lines). 
We fit a Gaussian to the peak of the cross-correlation, adopting 
the best-fit Gaussian centroid as the redshift for each source image.  
We  adopt the 
mean and sample standard deviation of the four measurements as 
the source redshift $z_{\rm S}=\zqso\pm\zqsoerr$.  
Observations at 
longer wavelengths %(covering, e.g., \ion{Mg}{2} or [\ion{O}{3}]
%emission; \citealt{vandenberk01}) 
are needed to constrain the redshift to higher precision.

\subsection{Intervening Absorption}\label{subsec.intervene}

%The KCWI spectra are of sufficient S/N and resolution to identify multiple intervening absorption-line systems.  
We visually inspected each  spectrum to identify foreground metal-line
absorption, focusing in particular on the identification of
\ion{C}{4} $\lambda\lambda 1548,1550$, 
\ion{Mg}{2} $\lambda\lambda 2796,2803$,
and \ion{Fe}{2} $\lambda\lambda 2586,2600$ transitions known to
arise in collisionally ionized or photoionized diffuse media at
temperatures $\sim10^{4-5}\,$K \citep{bergeronstasinska86}.  %Having
%identified these absorbers, 
%We interactively fit a spline function to the quasar continua %and used
%this fit 
%to generate continuum-normalized spectra.   
We fit a Gaussian
function to each absorber to determine its wavelength
centroid, and calculate a redshift for each absorption system (\zabs)
from the average redshift of the centroids of every associated
absorption feature.
We measure observed equivalent widths (\EWo) using a boxcar sum of the
flux decrement over the velocity range of each line, and compute rest
equivalent width using $\EWr=\EWo/(1+\zabs)$.

%We list these systems along with their redshifts and \EWr\ values in 
%Table \ref{tab.int}.  Each of these systems is also indicated with vertical colored lines in Figure~\ref{fig.kcwi1d}.
%we list intervening absorption systems identified in our visual inspection of the data. 
%Figure \ref{fig.abs} summarizes the values in Table \ref{tab.int}. 
%We interactively fit the quasar continua and normalize the spectra. 
We summarize these systems in Table~\ref{tab.int} and indicate them
with vertical lines in Figure~\ref{fig.kcwi1d}.
The reported uncertainties in the rest equivalent widths, \sigEWr\, do
not include errors associated with continuum normalization. The effect
of continuum-level placement on the measurement of equivalent width
can often be significant, especially %given the BAL nature of the
for BAL quasars observed at the 
medium spectral resolution of our KCWI configuration.  However, 
systematic error in continuum placement is likely to be similar across
the four sight lines.  We therefore expect
analysis of the relative variation in 
 \EWr\  at a given \zabs\ to be insensitive to these uncertainties.

%such that the results presented in Table \ref{tab.int} will still robustly reveal 
%systems in which the absorption strength varies significantly between
%the source images.  

%%JMO: the Abs_Sys.py code generates tab_abs.tex, copy/paste it here
%\onecolumngrid %% KLC: unnecessary... I just moved around to it looked nice-ish.

\startlongtable
\begin{deluxetable}{lccrr}
\tablewidth{0pc}
\tablecaption{Intervening Absorption-Line Systems\label{tab.int}}
\tablehead{
\colhead{Sight Line} &
\colhead{Redshift} &
\colhead{Transition ($\lambda_{\rm r}$)} &
\colhead{\EWr} &
\colhead{\sigEWr}
\\
 & & \colhead{\hfill(\AA)} & \colhead{(m\AA)} &\colhead{(m\AA)}
}
\startdata
A & 0.5775 & \ion{Mg}{2} (2796) & 96.2 & 4.6\\
A & 0.6069 & \ion{Mg}{2} (2796) & 437.2 & 3.9\\
A &  & \ion{Fe}{2} (2600) & 74.0 & 2.4\\
A & 1.0491 & \ion{Fe}{2} (2600) & 189.4 & 3.6\\
A & 1.7587 & \ion{C}{4} (1548) & 91.2 & 2.0\\
A & 1.7525 & \ion{C}{4} (1548) & 78.6 & 2.1\\
A & 2.0371 & \ion{C}{4} (1548) & 137.4 & 2.1\\
A &  & \ion{Si}{4} (1393) & 137.4 & 1.9\\
A &  & \ion{H}{1} (1216) & 1713.5 & 4.2\\
A & 2.0386 & \ion{C}{4} (1548) & 35.5 & 2.0\\
A & 2.1766 & \ion{C}{4} (1548) & 26.8 & 2.4\\
\hline
B & 0.5776 & \ion{Mg}{2} (2796) & 106.0 & 4.9\\
B & 0.6070 & \ion{Mg}{2} (2796) & 267.7 & 6.3\\
B &  & \ion{Fe}{2} (2600) & 22.8 & 2.2\\
B & 1.0491 & \ion{Fe}{2} (2600) & 71.5 & 4.1\\
B & 1.7586 & \ion{C}{4} (1548) & 100.2 & 2.2\\
B & 2.0371 & \ion{C}{4} (1548) & 111.0 & 2.3\\
B &  & \ion{Si}{4} (1393) & 111.0 & 2.0\\
B &  & \ion{H}{1} (1216) & 1741.2 & 4.4\\
B & 2.0388 & \ion{C}{4} (1548) & 28.03 & 2.3\\
B & 2.1767 & \ion{C}{4} (1548) & 24.7 & 2.4\\
\hline
C & 0.5772 & \ion{Mg}{2} (2796) & 143.6 & 7.5\\
C & 0.6068 & \ion{Mg}{2} (2796) & 376.9 & 6.1\\
C &  & \ion{Fe}{2} (2600) & 36.6 & 3.6\\
C & 1.0494 & \ion{Fe}{2} (2600) & 481.4 & 6.4\\
C & 1.7585 & \ion{C}{4} (1548) & 126.2 & 3.1\\
C & 2.0370 & \ion{C}{4} (1548) & 120.1 & 2.4\\
C &  & \ion{Si}{4} (1393) & 120.1 & 2.7\\
C &  & \ion{H}{1} (1216) & 1649.4 & 5.8\\
C & 2.0387 & \ion{C}{4} (1548) & 53.6 & 2.3\\
C & 2.1765 & \ion{C}{4} (1548) & 45.4 & 4.3\\
\hline
D & 0.5762 & \ion{Mg}{2} (2796) & 129.9 & 15.5\\
D & 0.5768 & \ion{Mg}{2} (2796) & 102.2 & 13.9\\
D & 0.6069 & \ion{Mg}{2} (2796) & 322.7 & 18.8\\
D &  & \ion{Fe}{2} (2600) & 42.8 & 10.8\\
D & 0.7584 & \ion{Mg}{2} (2796) & 965.5 & 20.2\\
D &  & \ion{Fe}{2} (2600) & 715.6 & 23.2\\
D & 1.0487 & \ion{Fe}{2} (2600) & 619.6 & 13.5\\
D & 1.7587 & \ion{C}{4} (1548) & 21.8 & 6.6\\
D & 2.0370 & \ion{C}{4} (1548) & 132.9 & 7.4\\
D &  & \ion{Si}{4} (1393) & 132.9 & 2.7\\
D &  & \ion{H}{1} (1216) & 1649.4 & 5.8\\
D & 2.0387 & \ion{C}{4} (1548) & 25.0 & 6.7\\
\enddata
\tablecomments{The observed wavelength of each of these absorbers is indicated above the corresponding source image spectrum in Figure~\ref{fig.kcwi1d}.}
\end{deluxetable}

Given the complexity of the spectra of \object\ and the resolution of these
data, we do not attempt to identify metal absorption lines in the
\Lyaf, and we are likely missing absorption lines contaminated by the
BAL features dominating the regions near the quasar emission lines. 
%(For example, the intrinsic \ion{C}{4} absorption extends from
%$\lambda_{\rm obs} \approx 4900$--5200\,\AA, covering
%$\approx\!10,000\kms$ blueward of the quasar emission; see Figure
%\ref{fig.kcwi1d}.)  
Higher resolution data will be required to perform a comprehensive
analysis of intervening absorption; however, we discuss some
preliminary findings based on the present dataset in Section
\ref{subsec.coherence}.

The KCWI data were not of sufficient depth to detect emission from the
lensing galaxy, and hence cannot directly constrain its redshift. 
%\cite{bergheaetal17ph} give a photometric estimate for the galaxy of $z_{\rm L} =  0.57^{+0.20}_{-0.13}$.  
We note that three intervening absorbers are found to have redshifts
within the $\pm1$-$\sigma$ photometric errors of the
\citeauthor{bergheaetal17ph} redshift estimate for the lens ($z_{\rm
  L}=0.57^{+0.20}_{-0.13}$); 
in addition, two of these systems are
  detected in all four source images. One of these latter systems,
  identified at $\zabs=0.5775$ in sight line A, has a redshift very
  close to the best-fit photometric estimate.
%Each absorption-line system has nearly exactly their claimed central value.  
Moreover, source image D (the image closest to the expected position of the
lens in projection) %; see Fig.\ 1 in \citealt{bergheaetal17ph})
exhibits a two-component absorbing structure in the system near
$z_{\rm L}=0.57$ (having $\zabs=0.5762$ and 0.5768).  These double
components are not exhibited in any of the other sight lines at
$\zabs=0.577$, and are suggestive of the complex, multiple-component
\ion{Mg}{2} absorbers typically observed close to bright galaxies
\citep[e.g.,][]{kacprzak10,chen14}.
Sight line D also exhibits 
%is also exceptional in that it has a 
strong \ion{Mg}{2}  absorption at $\zabs\approx0.76$ that is not detected in
the other sight lines due to blending with a higher-redshift
\ion{C}{4} absorber (at $\zabs=2.177$), raising the possibility that
this $\zabs\approx0.76$ system is associated with the lensing
galaxy. 
 %Moreover, unlike the other lensed images, D's intervening absorber nearest the value of $z_{\rm L}$ is seen to have two components, a structure not seen in the other   lensed sight lines.  
Without a spectrum of the lens itself, we cannot be certain
of its redshift; however, the detection of absorption in
every sight line at $\zabs\approx0.577$, its complex velocity
structure in sight line D, and the consistency of this redshift with
photometric constraints from
%complicated absorption-line system so close to the photometric redshift estimate of 
\citet{bergheaetal17ph} leads us to adopt $z_{\rm L}=\zabs=0.5768$ in our analysis below.

\subsection{Sight Line Geometry}\label{subsec.geometry}

To compute the transverse separation of the four sight lines as a function of absorber redshift, we refer to Equation (5) in \citet{cooke10}:
\begin{equation}
	S_0=\frac{\theta_{\rm obs}D_{\rm L}(D_{\rm S}-D_{\rm abs})}{(1+\zabs)(D_{\rm S}-D_{\rm L})}, 
\end{equation}
where $\theta_{\rm obs}$ is the observed angular separation between the sight lines and $D_X$ is the co-moving distance to the redshift $z_X$, with L, S, and `abs' indicating the lens, source, and absorber, respectively.  
%We note that in invoking this relation to compute the physical
%separation of any given sight line pair in a quadruple lens, 
For these calculations,
we assume that the source is located directly behind the center of mass of the lensing galaxy.

%\textbf{KHRR?: I hope this applies to a quad lens just as well as a double!  Seems like it should.}  
Adopting $z_{\rm L}=0.5768$ %the absorber redshift closest to the photometric redshift determined for the lens by \citet{bergheaetal17ph}---
and a source redshift  $z_{\rm S}=\zqso$, we show the resulting
physical separations between all sight-line combinations at the
redshifts of several of the systems identified in Section
\ref{subsec.intervene} in Table~\ref{tab.S0}.  If instead $z_{\rm
  L}=0.6069$ or  $z_{\rm L}=0.7584$, 
each distance would increase by a factor of $1.072$  or $1.476$,
respectively. 
The \object\ system thus permits assessment of the variation in the
strength and velocity structure of foreground absorption on scales of
a kiloparsec at $\zabs\approx 2$,  $\approx\!5$--10\,kpc at $\zabs
\approx1$, and $\approx\!10$--20\,kpc at $\zabs\approx0.6$.
%KHRR?: Can also discuss how sensitive this is to $z_{\rm L}$ (since it's only photometric for now).  
%Also, we don't have to keep the table, we could just summarize in the text since the %numbers are a bit repetitive.

%\onecolumngrid %% KLC: deluxetable* does this in emulateapj like figure* works

%% KLC: below is KHRR original hand-made table with lensing_model,/orig_cosm
% \begin{deluxetable*}{lcccccccc}
% \tablecaption{Projected Distances between Lensed Sight lines\label{tab.S0}}
% \tablehead{\colhead{Sight Line Pair} & \colhead{$\theta_0$} & 
% \colhead{$S_{0}\,(0.5768)$}  & \colhead{$S_{0}\,(0.6069)$} & 
% \colhead{$S_{0}\,(0.7584)$} & \colhead{$S_{0}\,(1.0491)$}  & 
% \colhead{$S_{0}\,(1.7525)$} & \colhead{$S_{0}\,(2.0371)$} & 
% \colhead{$S_{0}\,(2.1766)$}\\ 
%  & ($\arcsec$) & (kpc) & (kpc) & (kpc) & (kpc) & (kpc) & (kpc) & (kpc)} %% KLC: don't need \\
% \startdata
% A-B & 1.26  & 8.4  & 8.1 & 6.4 & 4.2  & 1.2 & 0.57 & 0.32\\
% A-C & 1.27 &  8.5 &  8.1 & 6.5 & 4.2  & 1.2 & 0.57 & 0.32\\
% A-D & 3.34 &  22.2 & 21.3 & 17.0 & 11.0 & 3.2 & 1.5 & 0.84\\
% B-C & 2.48 &  16.5 & 15.8 & 12.6 & 8.2  & 2.4 & 1.1 & 0.62\\
% B-D & 3.28 &  21.9 & 20.9 & 16.7 & 10.8 & 3.2 & 1.5 & 0.82\\
% C-D & 3.35 &  22.3 & 21.3 & 17.0 & 11.0 & 3.2 & 1.5 & 0.84 %% KLC: don't need final \\
% \enddata
% \tablecomments{All distances are computing assuming $z_{\rm L} = 0.5768$.  If instead $z_{\rm L} = 0.6069$, each distance is increased by a factor of $1.072$.}
% \end{deluxetable*}

%% KLC: WMAP5 (Komatsu et al.) cosmology
\begin{deluxetable*}{lcccccccc}
\tablecaption{Projected Distances between Lensed Sight Lines for $z_{\rm L} = 0.5768$\label{tab.S0}}
\tablehead{\colhead{Sight Line Pair} & \colhead{$\theta_0$} & 
\colhead{$S_0\,(0.5768)$} & \colhead{$S_0\,(0.6069)$} & \colhead{$S_0\,(0.7584)$} & \colhead{$S_0\,(1.0491)$} & \colhead{$S_0\,(1.7525)$} & \colhead{$S_0\,(2.0371)$} & \colhead{$S_0\,(2.1766)$}  \\
 & ($\arcsec$) & \colhead{(kpc)} & \colhead{(kpc)} & \colhead{(kpc)} & \colhead{(kpc)} & \colhead{(kpc)} & \colhead{(kpc)} & \colhead{(kpc)}  }
\startdata
A-B & 1.26 &  7.7 &  7.4 &  5.9 &  3.8 &  1.1 & 0.51 & 0.28 \\
A-C & 1.27 &  7.8 &  7.5 &  5.9 &  3.8 &  1.1 & 0.51 & 0.28 \\
A-D & 3.34 & 20.4 & 19.5 & 15.5 & 10.0 &  2.9 &  1.3 & 0.73 \\
B-C & 2.48 & 15.2 & 14.5 & 11.6 &  7.5 &  2.2 &  1.0 & 0.54 \\
B-D & 3.28 & 20.1 & 19.2 & 15.3 & 9.9 &  2.9 &  1.3 & 0.72 \\
C-D & 3.35 & 20.5 & 19.6 & 15.6 & 10.1 &  2.9 &  1.3 & 0.73 \\
\enddata
%\tablecomments{All distances are computing assuming $z_{\rm L} = 0.5768$.  If instead $z_{\rm L} = 0.6069$, each distance would increased by a factor of $1.072$.}
\end{deluxetable*}

%\twocolumngrid

\begin{figure}[hbt]
 \begin{center}
  \includegraphics[width=0.5\textwidth,clip=true,trim=0.6cm 0.5cm -0.3cm 0]{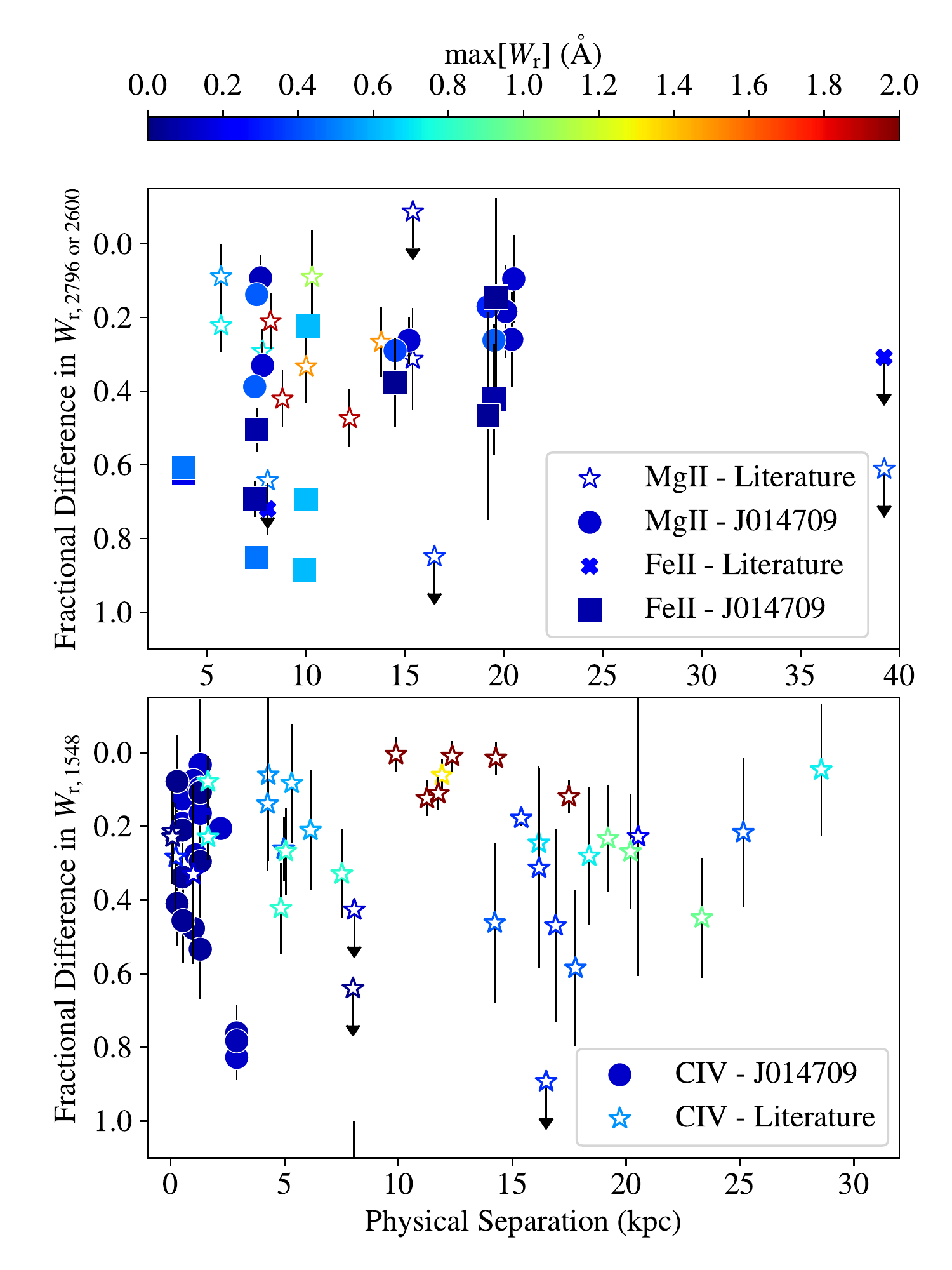}
  \end{center}
  \caption{Fractional difference in \EWr\ as a function of sight-line
    separation at \zabs\ for intervening absorbers detected toward
    \object\ (filled circles and squares) and toward lensed QSOs
    analyzed in previous studies (stars and crosses; 
\citealt{young81,foltz84,smette92,smette95,monier98,lopez99,chen14}).
The fractional difference is 
$(\EWlin{ion}^{\rm X}-\EWlin{ion}^{\rm Y})/\EWlin{ion}^{\rm X}$, where sight line X has the stronger
absorption of the two systems. 
 Points are color-coded to indicate \EWlin{X}, and the vertical scale
 increases  top-to-bottom (i.e., more-to-less coherent). 
The upper panel shows \EWr\ offsets in \ion{Mg}{2} $2796$
(filled circles) and 
\ion{Fe}{2} $2600$ (filled squares).  
The lower panel shows the same measurements for \ion{C}{4} $1548$.
%Crosses and stars indicate \EWr\ offsets collected from several
%previous works analyzing high-S/N spectroscopy of lensed quasars
%\citep{young81,foltz84,smette92,smette95,monier98,lopez99,chen14}.  
}
 \label{fig.abs}

\end{figure}

%\subsection{Relation between \EWr\ Across Multiple, Close Sight Lines}
\subsection{Coherence in \EWr\ across Multiple, Close Sight Lines}\label{subsec.coherence} 
%% KLC: suggestion; also falls on line

The analysis described in Section \ref{subsec.intervene} revealed
 7--8 securely-detected 
intervening absorption systems toward every quasar image.  For all of
these systems, with the exception of the \ion{C}{4} absorber at
$\zabs=1.7525$ in sight line A and the \ion{Mg}{2} absorber at
$\zabs=0.7584$ in sight line D,
%\footnote{The strong $\zabs = 0.7584$ \ion{Mg}{2} lines contaminate the weaker, $\zabs = 2.177$ \ion{C}{4} system detected in sight lines A--C.}
we detect counterpart absorbers within a velocity range $\delta v\pm250\kms$, 
with the vast majority of the counterparts lying within $\delta
v\pm100\kms$.  Given the physical separation of the sight lines, this
finding is suggestive of absorbing structures extending over
relatively large scales (e.g., $>\!5$--20\,kpc at $\zabs<1$).  
To quantitatively assess the physical extent of these absorbers and
the scale over which their \EWr\ varies, we compute the fractional
difference in \EWr\ values measured at a given \zabs\ for each pair of
sight lines, 
$(\EWlin{ion}^{\rm X}-\EWlin{ion}^{\rm Y})/\EWlin{ion}^{\rm X}$, where
sight line X has the stronger absorption of the two.  We show these
fractional differences for \ion{Mg}{2} (filled circles) and
\ion{Fe}{2} (filled squares) systems in the upper panel of Figure
\ref{fig.abs}
vs.\ physical separation of the sight line pair (taken from Table
\ref{tab.S0}).  The lower panel of the Figure shows the same
measurements for our \ion{C}{4} systems (filled circles).  Each point
is color coded by the corresponding value of $\EWlin{ion}^{\rm X}$.
 %\footnote{Fractional \EWr\ differences for the $\zabs = 0.7584$ \ion{Mg}{2} system in sight line D and for the 
%The strong $\zabs = 0.7584$ \ion{Mg}{2} lines contaminate the weaker, $\zabs = 2.177$ \ion{C}{4} system detected in sight lines A--C.}
%displays the fractional difference as a function of the physical separation, using the values in Table \ref{tab.S0}; fractional difference is $(\EWlin{ion}^{\rm X} - \EWlin{ion}^{\rm Y})/\EWlin{ion}^{\rm X}$, where sight line X has the stronger absorption of the two. 

Considering the two \ion{Mg}{2} systems at $\zabs\approx0.577$ and
$0.607$, we measure small fractional $\EWlin{2796}$ differences
($<40\%$) across the full range of sight line separations
($\sim7-21\,$kpc), pointing to a high degree of coherence over large
scales even for these relatively weak absorbers (having
$\EWlin{2796}\approx0.1-0.44\,$\AA). 
The two \ion{Fe}{2} systems, on the other
hand, exhibit a much greater degree of variation in absorption
strength, yielding fractional $\EWlin{2600}$ differences of
$\approx15-90\%$ between every sight line pair. We find that the four
\ion{C}{4} systems exhibit a quite high degree of variation, with
$\EWlin{1548}$ differences ranging between $\approx5$ and $80\%$ over
$0.3-3\,$kpc separations.  Such a wide range of $\EWlin{1548}$ values
points to 
gas densities which vary on sub-kpc scales
for the weakest of these absorbers (with $\EWlin{1548}\sim0.02-0.05\,$\AA).

%This analysis reveals that 
%the absorption-line strengths of many intervening systems vary significantly ($>\!50\%$) across sight lines (e.g., the \ion{Fe}{2} systems at $\zabs\approx 1.05$ and separated by $\approx\!10\,$kpc,  and the \ion{C}{4} systems at $\zabs\approx 1.76$ separated by only $\approx\!1$--3\,kpc), indicative of varying gas densities or velocity structure across the extent of the absorber.  Indeed, the $\zabs\approx 0.76$ absorber is both strong and present only in sight line D. 

%% KLC: since \lambda\lambda 1548,1550 used in abstract and earlier, no need to continue notation here
Figure \ref{fig.abs} also includes similar measurements collected from
the literature analyzing intervening \ion{C}{4} 1548, \ion{Mg}{2}
2796, and \ion{Fe}{2} 2600 absorption along lensed quasar sight lines.
For each system, we adopt updated constraints on $z_{\rm L}$ where
available and recalculate the transverse physical sight-line
separations assuming the WMAP5 \citep{komatsuetal09} cosmology (with
one exception described below).
%from  the literature.  Starting with \citet{young81}, an early work analyzing \ion{C}{4} absorption equivalent widths in spectroscopy of a doubly-lensed QSO, we have attempted to include every available \ion{C}{4} 1548, \ion{Mg}{2} 2796, and \ion{Fe}{2} 2600 equivalent width measurement from articles identifying such systems along lensed quasar sight lines.  
In detail, we include %\ion{C}{4}, \ion{
systems from
\citet{young81,foltz84,smette92,smette95,monier98,lopez99} and
\citet{chen14}.
For absorbers from \citet{foltz84}, we assume $z_{\rm L}=1.49$ and
an angular sight-line separation of $7.13\arcsec$ \citep{sol84}.  
In the case of absorbers from \citet{smette95}, we assume 
$z_{\rm  L}=1$ as in \citet{lopez99}.  The \citet{monier98} study focused on
the Cloverleaf lens, for which $z_{\rm L}$ is still not precisely
known; here we adopt $z_{\rm L}=1.88$ as estimated in
\citet{goicoechea10}. 
For absorbers from \citet{chen14}, we do not compute sight-line separations, and instead
adopt the physical distances listed in that work (although they assume
slightly different cosmological parameters).
Fractional \EWr\ differences (or lower limits in
cases in which an absorber is securely detected along only one sight
line) for each of these %\ion{C}{4} 
systems are indicated with open
stars (\ion{C}{4} and \ion{Mg}{2}) and filled crosses (\ion{Fe}{2}) in Figure~\ref{fig.abs}.
\footnote{There are a few additional studies
  which offer detailed analysis of metal-line absorption along lensed
  quasar sight lines \citep[e.g.,][]{rauch99,rauch02,churchill03};
  however, because these authors did not report equivalent widths, we
  do not include them here.}

%We also include the \ion{Mg}{2} and \ion{Fe}{2} absorbers 
%reported in a subset of these studies
%\citep{smette95,monier98,lopez99}.  %, adopting the same adjustments to
%$z_{\rm L}$ and sight line  separation described above.  
%We include $\EWlin{2796}$ measured in two systems known to arise
%within projected distances $\lesssim50\,$kpc
%of two luminous foreground galaxies along the sightlines of the
%quadruply lensed QSO HE 0435-1223 as discussed in \citet{chen14}.  In
%this case, we do not compute sight-line separations, and instead
%adopt the physical distances listed in that work (although they assume
%slightly different cosmological parameters). The fractional \EWr\
%differences for all of these systems are shown in the upper panel of
%Figure~\ref{fig.abs} with filled stars (\ion{Mg}{2}) and filled
%crosses (\ion{Fe}{2}).   

%While the number of data points is small, 
%Figure \ref{fig.abs} [WORKING HERE]
%Broadly speaking, Figure~\ref{fig.abs} 
Taken together, these measurements point to varying degrees of spatial
coherence for the three metal-line transitions examined.  
In the case of \ion{Mg}{2} $\lambda2796$, 22 of the 25 absorber pairs
exhibit fractional $\EWlin{2796}$ differences of $<0.5$
over a wide range of sight line separations ($6-21\,$kpc).  This
level of coherence is exhibited by both strong
($\EWlin{2796}>1.0\,$\AA) 
and weak  ($\EWlin{2796}<0.3\,$\AA) systems.  
Eight of these pairs %(or $\sim 21\%$ of the sample) 
have fractional differences $<0.2$, and again have a wide range of
maximum $\EWlin{2796}$.  Such small \EWr\ variations are suggestive of
\ion{Mg}{2}-absorbing clouds or cloud complexes extending over areas
several kiloparsecs across for both strong and weak \ion{Mg}{2} systems.

Turning to the \ion{Fe}{2} absorbers, we find that the majority of their fractional
\EWr\ differences are  $>0.5$.  While the number
of sight lines probing this transition is small, this suggests that
%the coherence length 
the bulk of the 
\ion{Fe}{2}-absorbing material in these structures % structures are typically
typically extends over
$\lesssim5\,$kpc.  Additional measurements probing smaller physical
separations will be valuable for tightening this constraint.

Finally, we find that the \ion{C}{4} $\lambda1548$ systems  exhibit a
high degree of coherence over the physical scales probed. 
%(which are
%much smaller than the physical separations of the sight lines probing
%the low ions discussed above).  
Among the 57 absorber pairs
shown, 50 have fractional $\EWlin{1548}$ differences $<0.5$, and 22
have fractional differences $<0.2$.  Generally speaking, small
fractional differences are exhibited over the full range of maximum
$\EWlin{1548}$ values; however, we also note that the strongest
absorbers in the sample (with $\EWlin{1548}>1.2\,$\AA) all have
fractional differences $\lesssim0.15$.  Again, this finding points to
\ion{C}{4}-absorbing structures extending over at least $\sim1-30\,$kpc,
and is suggestive of longer coherence lengths for stronger absorbers. 
 %A larger sample size is needed to verify such a scenario.

%% KHRR: have moved some of this to discussion above.
%However, we note from Figure \ref{fig.abs} that the absorption-line strengths of many intervening systems vary across sight lines (e.g., the systems at $\zabs\approx 1.05$ and $\zabs\approx 1.76$), indicative of varying gas densities across the extent of the absorbing gaseous structure.  Indeed, the $\zabs\approx 0.76$ absorber is both strong and present only in sight line D.  In addition to \ion{C}{4} and \ion{Si}{4} absorption, the $\zabs \approx 2.04$ system exhibits strong \ion{H}{1} absorption, indicative of a Lyman-limit system (Those \ion{H}{1} absorbers with column density $\lnhi > 17.2$).  We intend to explore these results in a future work with higher-resolution spectroscopy covering a wider wavelength range. 

\section{Discussion}\label{sec.discuss}

%\subsection{Outline of Literature on Intervening Absorption toward Lensed QSOs\label{subsec.igmprobes}}
The study of intervening absorption-line systems observed in
 background quasar spectroscopy has proven essential for assessing,
e.g., the neutral gas content of the Universe \citep{wolfe05} and 
the evolution of its metal content \citep{simcoe11,lehner14}.
%and the mass of the diffuse baryons in regions extending $>100\,$ kpc from luminous galaxies \citep{werk14, prochaska17}.
In spite of these advances,  numerous open questions remain
regarding the physical nature of the absorbers themselves.
Cosmological simulations predict that many of these systems arise in
the environments of luminous galaxies, tracing cool inflowing
streams or large-scale outflows driven by star formation
\citep[e.g.,][]{fumagalli11,shen13,faucher-giguere15}.  However, it
has proven difficult to leverage such predictions for constraints on
the physical origins of a given observed absorber population.

A crucial limitation has been our lack of information on the sizes and
morphologies of the absorbing structures.  Because background quasars
provide only a pencil-beam probe, constraints on the physical extent
of these systems have been obtained via 
modeling of the ionization state of the gas
\citep{churchillcharlton99,stocke13,werk14}.
However, these analyses are subject to substantial systematic
uncertainties (e.g., in the shape of the extragalactic ionizing
background spectrum, %in the intrinsic metal abundance pattern, 
and in the cloud geometry / density profile), and typically may only be used
to constrain the cloud thickness to within an order of magnitude
\citep[e.g.,][]{werk14}.
%(with, e.g., Cloudy photoionization models; \citealt{ferland98}).  

%In principle, the thickness of an absorbing cloud is manifest in the ratios of column densities arising from 
 
%Despite decades  of study and  significant advances in both observational and theoretical techniques, some of the most basic questions about absorption-line systems remain un-answered:  What are their sizes?  What galaxy types are they associated with? Do they represent inflowing, outflowing, or recycled gas?  

%\textbf{KLC: I enforced $h^{-1}$ for sizes from various other results b/c then the reader has to go do the conversion; technically we should just do the math to make it all fairly comparable. For example, Rauch et al. (1999) used $h = 0.5$ so that's a factor of two.} 
%The situation is particularly difficult at $\zabs>1$ where cosmological dimming often makes the galaxy hosts of absorbers beyond the reach of even the largest telescopes.  

Spectroscopy of multiple, close background sight lines offers a
valuable alternative probe of absorber morphology by mapping the
transverse dimension.  Gravitationally-lensed quasars are perhaps the
most efficient such sources, as they may be very bright, and produce
similar continua at a given $\lambda_{\rm obs}$. 
%(unlike projected  quasar pairs; e.g., \citealt{rubin15}). 
% This technique was introduced over three decades ago 
%\citep{young81,weymannfoltz83, foltz84}, and 
%Many of the relevant measurements from studies using this technique are summarized in Figure~\ref{fig.abs}.
Overall, the \EWr\ variations observed in these systems (summarized in
Figure~\ref{fig.abs})
indicate that either (1) each gas cloud composing the absorbers
extends across the physical separation of the beams; or (2) they arise
from extended structures made up of numerous smaller clouds with
similar velocity spread and/or column density along any given sight line.
 %(and with a high covering fraction).  

Ultimately, higher spectral resolution ($\mathcal{R}>6000$) will be
required to distinguish between these scenarios, as it permits
detailed comparison of the column densities, velocity centroids, and
line widths of  individual absorbing components across the sightlines
\citep[e.g.,][]{rauch99,rauch01,rauch02,chen14}.  A spectroscopic
survey for galaxies associated with these absorbers will allow us to
establish their context within the CGM, 
%and will additionally provide an alternative 
and will in addition test a basic
assumption invoked by most CGM studies to date. 
\ion{Mg}{2} absorbers with strengths similar to those in
our sample (with $\EWlin{2796}\approx0.1-1\,$\AA) are common
within projected distances $R_{\perp}<50\,$kpc of $\sim L^*$
galaxies at low redshift ($z\sim0.2$; \citealt{chen10}). 
However, constraints on the $\EWlin{2796}$ distribution in these
environments come from the assembly of numerous projected QSO-galaxy
pairs, each of which probes an independent halo. 
%Analysis of this distribution has revealed an intrinsic scatter in $\log \EWlin{2796}$ of $\sigma_{\rm C} = 0.196$ at a given $R_{\perp}$ and host galaxy absolute magnitude ($M_{\rm B}$), implying a fractional $\EWlin{2796}$ difference of $\approx 0.6$ (between values of $\log \EWlin{2796} \pm \sigma_{\rm C}$; \citealt{chen10}).  This level of variation is slightly higher than that exhibited by most of the \ion{Mg}{2} systems examined here, suggesting that the variation in $\EWlin{2796}$ from halo to halo may be larger than the small-scale variation at a given $R_{\perp}$ in a particular system. 
Larger samples of lensed QSOs probing foreground systems
\citep[e.g.,][]{chen14} will provide a critical test of the standard
interpretation that the \EWr\ distribution of absorbers observed toward
numerous QSO-galaxy pairs is representative of the absorption profile
in an individual galaxy CGM.

Moreover, measurement of the sizes,  velocity coherence, and metallicity of such
absorbers 
%(along with limits on their metallicity) 
will permit quantitative comparisons to the CGM features predicted in hydrodynamical `zoom' simulations, differentiating between smooth accretion streams, gas associated with infalling satellites, and the turbulent, clumpy flows arising from stellar feedback \citep[e.g][]{shen13,nelson15,faucher-giguere15,fielding17}.  Such constraints will in addition be necessary to test formalisms suggesting that the CGM cools via ``shattering'', resulting in a high covering fraction of parsec-scale, photoionized cloudlets \citep{mccourt16}. 
With $>2000$ lensed QSOs expected to be discovered in the ongoing PS1
and Dark Energy Surveys \citep{des16}, 
%,ostrovski17,lin17}, 
and several thousand to be uncovered by LSST \citep{ivezic08,oguri10}, this technique will soon become the state of the art in CGM studies.

\acknowledgements 
KLC acknowledges support from NSF grant AST-1615296. %\textbf{KLC:
                               % KHRR, don't forget to add mahalos to
                               % funding, as applicable.}

The data presented herein were obtained at the W.~M.~Keck Observatory,
which is operated as a scientific partnership among the California
Institute of Technology, the University of California and the National
Aeronautics and Space Administration. The Observatory was made
possible by the generous financial support of the W.~M.~Keck
Foundation.

The authors wish to recognize %and acknowledge 
the very significant
cultural role and reverence that the summit of Maunakea has always had
within the indigenous Hawaiian community.  We are most fortunate to
have the opportunity to conduct observations from this mountain.

Finally, the authors dedicate this work to the memory of Jerry Nelson, without whom the immense contributions to science made by the Keck Observatory would not be possible.

\facility{Keck:II (KCWI)}

%\clearpage
%%%%%%%%%%%%%%
%% BIBLI %%
%%%%%%%%%%%%%%
\bibliographystyle{apj}
%\bibliography{quad.bib}

\end{document}